\documentclass[12pt]{article}
\usepackage{amsmath}
\usepackage{amsfonts}
\usepackage{graphicx}
\usepackage{mathrsfs}

\begin{document}

%
\def\papertitlepage{\baselineskip 3.5ex \thispagestyle{empty}}
\def\preprinumber#1#2{\hfill \begin{minipage}{4.2cm}  #1
                 \par\noindent #2 \end{minipage}}
\renewcommand{\thefootnote}{\fnsymbol{footnote}}
\newcommand{\beq}{\begin{equation}}
\newcommand{\eeq}{\end{equation}}
\newcommand{\beqa}{\begin{eqnarray}}
\newcommand{\eeqa}{\end{eqnarray}}
\newcommand{\n}{\nonumber\\}
\catcode`\@=11
\@addtoreset{equation}{section}
\def\theequation{\thesection.\arabic{equation}}
\catcode`@=12
\relax
%
%
%
\papertitlepage
\setcounter{page}{0}
\preprinumber{KEK-TH-1474}{}
\baselineskip 0.8cm
\vspace*{2.0cm}
\begin{center}
{\large\bf Taub-NUT Crystal}
\end{center}
\vskip 4ex
\baselineskip 1.0cm
\begin{center}
           {Harunobu Imazato${}^\sharp$, Shun'ya Mizoguchi${}^{\sharp~\!\!\flat}$ 
           and Masaya Yata${}^\sharp$} 
\\
\vskip 1em
     ${}^\sharp$  {\it 
Department of Particle and Nuclear Physics}\\
 \vskip -2ex {\it The Graduate University for Advanced Studies}\\
 \vskip -2ex {\it Tsukuba, Ibaraki 305-0801, Japan} \\
\vskip 1em
      ${}^\flat$ {\it Theory Center}\\
 \vskip -2ex {\it High Energy Accelerator Research Organization (KEK)} \\
       \vskip -2ex {\it Tsukuba, Ibaraki 305-0801, Japan} \\
\end{center}
\vskip 5ex
%
\baselineskip=3.5ex
\begin{center} {\bf Abstract} \end{center}

\vskip 2ex
We consider the Gibbons-Hawking metric for a three-dimensional periodic array 
of multi-Taub-NUT centers, containing not only centers with a positive 
NUT charge but also ones with a negative NUT charge. The latter are regarded 
as representing the asymptotic form of the Atiyah-Hitchin metric. 
The periodic arrays of Taub-NUT centers have close parallels with ionic crystals, 
where the Gibbons-Hawking potential plays the role of the Coulomb static potential 
of the ions, and are similarly classified according to their space groups.
After a periodic identification and a ${\bf Z}_2$ projection, the array is 
transformed by T-duality to a system of NS5-branes with the SU(2) 
structure, and a further standard embedding yields, though singular, a half-BPS 
heterotic 5-brane background with warped compact transverse dimensions.
A discussion is given of  the possibility of probing the singular geometry by 
two-dimensional gauge theories.  


\vspace*{\fill}
\noindent
July 2011
\newpage
\renewcommand{\thefootnote}{\arabic{footnote}}
\setcounter{footnote}{0}
\setcounter{section}{0}
\baselineskip = 0.6cm
\pagestyle{plain}

\section{Introduction}
\indent 
Branes and singularities are essential ingredients in modern string theory. 
On one hand, branes are non-perturbative objects whose existence is required  
for consistency of various dualities. 
On the other hand, singularities are sources from which the effects of 
such objects emerge by allowing them to wrap around collapsing cycles. 
Branes and/or singularities are also required 
in order to realize inflation in string theory; under some mild assumptions, 
it was shown that  smooth compactifications cannot 
allow an accelerating solution  \cite{no-go_theorem}. 
One way to circumvent this is to consider warped compactification with branes
or singularities. There is a growing literature on attempts to realize a Randall-Sundrum-like 
framework \cite{RS} in string theory by using D-branes \cite{V,GKP,KKLT}. 

It has been known that, 
in order to achieve warped compactification in supergravity, 
some of the constituent branes need to have negative tension \cite{GKP}.
(This fact was shown in type IIB theory but can be generalized to other 
string theories.)
In type II theories, an orientifold 
has 
this property. 
In {\it heterotic} string theory, however,  
one cannot define orientifolds  
as one does in type II theories, since, 
while one needs to gauge world-sheet parity to define orientifolds, 
heterotic theory is left-right asymmetric and hence does not possess such a 
symmetry to be gauged. Although even in the absence of the worldsheet definition 
one can define orientifolds by simply assigning ${\bf Z}_2$ actions on various 
fields \cite{HananyKol}, it is unclear what kind of geometry is produced there.    
The back-reaction problem of orientifolds was considered in \cite{backreaction}.

In this paper, we propose a novel, interesting set up for warped 
compactification of heterotic (as well as type II) string 
theory to six dimensions, based on a simple idea: 
We consider the Gibbons-Hawking metric  \cite{GH} for a three-dimensional 
periodic array of multi-Taub-NUT centers consisting of not only  centers 
with a positive NUT charge but also  ones having a negative NUT charge. 
This negative-charge metric is regarded as representing the 
asymptotic form of the Atiyah-Hitchin metric; as is well known, 
the Atiyah-Hitchin metric \cite{AH} is a smooth self-dual metric on a Euclidean 
four-manifold, and it looks like a negative-charge Taub-NUT for an 
observer who is far enough from the bolt locus. 
Such a periodic 
array of infinite Taub-NUT centers was already discussed in \cite{CGLP,KatieR}.
The idea of introducing an Atiyah-Hitchin space in the Gibbons-Hawking 
metric was used in \cite{Sen} to understand how an $SO(2n)$ gauge 
symmetry enhancement occurs due to D-branes.
By T-duality, 
the periodic array of Taub-NUT centers is converted to a periodic array of 
NS5-branes. An ordinary
(positive-charge) Taub-NUT is T-dualized to a (smeared) NS5-brane, and
the negative-charge Taub-NUT is mapped to 
a (smeared) negative-tension brane. 
The resulting T-dual system has the $SU(2)$ structure (hyper-K\"{a}hler 
with torsion),
and preserves one half  the supersymmetries. We then embed the 
generalized spin connection into gauge connection to obtain 
a 5-brane system with a compact four-dimensional transverse space 
in {\it heterotic} string theory.

As is known, although the
negative-NUT approximation of the Atiyah-Hitchin metric is 
good at large distances compared to the NUT-charge scale, 
near the center the approximate metric 
becomes singular on a two-dimensional sphere (in the three-dimensional 
space where the periodic array is defined) of radius equal to the 
NUT charge. Therefore, on the T-dual side, there are two kinds of regions 
where the supergravity analysis is not reliable: One is the region 
near a smeared NS5-brane, the familiar one where the geometry 
is an infinite throat and the dilaton blows up as one approaches the core.  
The other is the region near the brane dual to the nagative-charge Taub-NUT center, 
in which, unlike the above, both the metric and the dilaton go to zero 
as one approaches the singularity surface, and inside it the metric becomes 
negative and the space ceases to be Riemannian. The former is a common 
difficulty, while the latter singularity is the one inherited from the ``one-loop 
approximation" of the Atiyah-Hitchin metric without taking  into 
account the instanton effects.
Nevertheless, despite the existence of these regions, we would like to emphasize 
that our construction is interesting in the following points:

First, it clarifies the origin of orientifold-like objects in heterotic string theory;
such objects are necessary to be included to neutralize the total NUT 
charge before T-duality. Indeed, their T-duals are very similar to orientifolds; 
the ${\bf Z}_2$ projection naturally arises from the bolt of the Atiyah-Hitchin space, and 
the ADM energy per unit area is negative. 
The second point we wish to make is that our 5-brane system is not just 
being a singular geometry, but suggests how the nonperturbative effects 
(by vortices in the corresponding gauge theory) correct the geometry.
Although incomplete, we will consider in a later section the relevant 
$(4,4)$ sigma models whose leading-order moduli spaces are singular 
5-brane geometries above but vortex corrections may change them.  
The problem amounts to finding a possible deformation of 
hyper-K\"{a}hler geometry with torsion, and the sigma model  \cite{GHR} analysis 
may provide a systematic way.

The infinite periodic array of Taub-NUT centers that we describe is 
similar to three-\! dimensional ionic crystals, such as sodium 
chloride (NaCl), cesium chloride (CsCl), etc., where the Atiyah-Hitchin's 
and Taub-NUT's correspond to minus and plus ions in the crystal, 
and the Gibbons-Hawking potential plays the role of the Coulomb 
static potential of the ions. 
Generically, the Gibbons-Hawking potential for such a
three-dimensional periodic 
array does not converge, even if the total NUT charge of a unit cell is 
neutralized. As we will see shortly, the condition for convergence is that 
the positive- and negative-charge centers are arranged in a unit cell 
in such a way that the leading difference operator of the multipole expansion 
of the potential becomes proportional to the discrete Laplace operator. 
This condition constrains the locations of the Taub-NUT centers relative 
to the  Atiyah-Hitchin centers, that is, the ones with  negative NUT charge. 
Another constraint for the positions of the Taub-NUT centers is that 
they must be inversion symmetric with respect to every 
Atiyah-Hitchin bolt locus so that they are consistent with the ${\bf Z}_2$ 
identification around the bolt.  Thus, while a real crystal is mathematically 
described as a three-dimensional torus, the ``Taub-NUT crystal" is a certain   
orbifold.

The plan of this paper is as follows: In section 2 we review the basic facts 
about the Atiyah-Hitchin metric. In section 3 we give a definition of our 
set up by introducing a three-dimensional periodic array of Taub-NUT centers, 
and examine the convergence condition for the Gibbons-Hawking potential. 
A relation to the $K_3$ orientifold is discussed, and the classification of the 
crystal structure is also described there. In section 4 we take T-duality of 
the Taub-NUT crystal to obtain a 5-brane system with compact transverse 
dimensions, and then we embed it into heterotic string theory. At the end 
of this section we briefly discuss possible corrections to the singularity from 
the point of view of the nonlinear sigma model approach. Section 5 is devoted 
to a summary. An appendix contains some details of the Darboux-Halphen 
system, where comparisons are also made between the solutions in the literature.

\section{The Atiyah-Hitchin metric}
Let us first recall some basic properties of the Atiyah-Hitchin metric.
The Atiyah-Hitchin metric is a smooth metric on a Euclidean four-manifold with a(n) (anti-)self-dual 
Riemann curvature without triholomorphic $U(1)$ isometry. It is  written 
as a special Bianichi IX metric of the form
\beqa
ds^2= a^2 b^2 c^2 dt^2 + a^2 \sigma_1^2 + b^2 \sigma_2^2 + c^2 \sigma_3^2,
\label{BianchiIX}
\eeqa
where $a$, $b$ and $c$ are functions of $t$ only.  $\sigma_i$'s $(i=1,2,3)$ are 
the Maurer-Cartan 1-forms of $SU(2)$ given by
\beqa
\sigma_1&=& -\sin\psi d\theta + \sin\theta \cos\psi d\phi,\n
\sigma_2&=& \cos\psi d\theta + \sin\theta \sin\psi d\phi,\\
\sigma_3&=& d\psi + \cos\theta d\phi.\nonumber
\eeqa
In terms of a new set of functions
\beqa
w_1=bc,~ w_2=ca,~ w_3=ab, 
\eeqa 
a (sufficient) condition for a(n) (anti-)self curvature is that they fulfill the 
Darboux-Halphen system
\beqa
\dot{w}_1 + \dot{w}_2 = 2 w_1  w_2, \n
\dot{w}_2 + \dot{w}_3 = 2 w_2  w_3, 
\\
\dot{w}_3 + \dot{w}_1 = 2 w_3  w_1. \nonumber  
\eeqa
For our purpose, a convenient representation of solutions is 
the one in terms of elliptic theta functions \cite{Ohyama, HP}
\beqa
w_j(t)&=&2\frac{d}{dt}\log\vartheta_{j+1}(0,i\alpha t)~~~(j=1,2,3),
\label{wjsection2}
\eeqa
where $\alpha>0$ is an arbitrary constant parameter. We will see 
that this $\alpha$ is related to the NUT charge of the asymptotic Taub-NUT 
metric. 
The obtained self-dual metric is thus
\beqa
ds^2&= &2\left(\prod_{j=3,4,2}\frac{d}{dt}\log\vartheta_{j}(0,i\alpha t)
\right)
 \left( 4 dt^2 + \frac{\sigma_1^2
}{(\frac{d}{dt}\log\vartheta_{2}(0,i\alpha t))^2
} \right. \n
&&~~~~~~~~~~~~~~~~~~~~~~~~~~~~~~~~~~~~~\left.+ \frac{\sigma_2^2 
}{(\frac{d}{dt}\log\vartheta_{3}(0,i\alpha t))^2
} 
+ \frac{\sigma_3^2
}{(\frac{d}{dt}\log\vartheta_{4}(0,i\alpha t))^2
}\right). 
\label{BianchiIXAH}
\eeqa
In this representation of the solutions, Atiyah-Hitchin's original solution 
corresponds to $\alpha=\pi$, whereas Gibbons-Manton's solution to  $\alpha=2\pi$ 
(with replacing the radial coordinate $t$ by $-\eta$. See Appendix for more detail.).

It is well known \cite{GM} (also see \cite{YW} for a review) that, 
on one hand, the Atiyah-Hitchin metric develops 
a removable ``bolt"  singularity at the origin (in a certain coordinate 
system), and on the other hand, it rapidly approaches to a 
Taub-NUT metric with  negative NUT charge asymptotically. 
Although these are well-known facts, we include a brief derivation 
of them because they are essential in constructing a crystal structure 
in the space of Taub-NUT centers.  
These behaviors can be
conveniently confirmed in our representation of solutions in terms of theta 
functions \cite{HP}.

\noindent{\it (i) The behavior near $t=\infty$: The bolt}

The elliptic theta functions are readily given as an infinite series as
\beqa
\vartheta_3(0,\tau)&=&\sum_{n\in{\bf Z}}q^{\frac12 n^2},\n
\vartheta_4(0,\tau)&=&\sum_{n\in{\bf Z}}(-1)^n q^{\frac12 n^2},\\
\vartheta_2(0,\tau)&=&\sum_{n\in{\bf Z}}q^{\frac12 (n-\frac12)^2},\nonumber
\eeqa
with $q=e^{2\pi i \tau}$. Putting $\tau=i\alpha t$, 
the $t\rightarrow\infty$ limit of (\ref{wjsection2}) can be obtained by 
taking the leading terms of the expansions:
\beqa
w_2(t)&\stackrel{t\rightarrow\infty}{\sim}&2\frac d{dt}\log(1+ 2 e^{-\pi  \alpha t}+\cdots)\n
&=&-4\pi \alpha ~e^{-\pi\alpha t}+\cdots,\n
w_3(t)&\stackrel{t\rightarrow\infty}{\sim}&2\frac d{dt}\log(1- 2 e^{-\pi  \alpha t}+\cdots)\n
&=&+4\pi \alpha ~e^{-\pi\alpha t}+\cdots,\\
w_1(t)&\stackrel{t\rightarrow\infty}{\sim}&2\frac d{dt}\log(2 e^{-\frac{\pi}4  \alpha t}+\cdots)\n
&=&-\frac\pi 2 \alpha+\cdots.\nonumber
\eeqa 
Therefore, $a$,$b$ and $c$ are
\beqa
a^2&=&\frac{w_2 w_3}{w_1}
~\stackrel{t\rightarrow\infty}{\sim}~
32\pi \alpha e^{-2 \pi \alpha t}+\cdots,\n
b^2&=&\frac{w_3 w_1}{w_2}
~\stackrel{t\rightarrow\infty}{\sim}~
\frac\pi 2 \alpha+\cdots,\n
c^2&=&\frac{w_1 w_2}{w_3}
~\stackrel{t\rightarrow\infty}{\sim}~
\frac\pi 2 \alpha+\cdots.\nonumber
\eeqa
The metric takes the form
\beqa
ds^2&\stackrel{t\rightarrow\infty}{\sim}&
8\pi\alpha(\pi^2\alpha^2 e^{-2\pi\alpha t} dt^2
+4 e^{-2\pi\alpha t }\sigma_1^2)
+\frac\pi 2 \alpha(\sigma_2^2 + \sigma_3^2).
\eeqa
This is a bolt; by a change of coordinates $u\equiv e^{-\pi\alpha t}$, the metric becomes
\beqa
ds^2&=&8\pi\alpha(du^2 + 4u^2 \sigma_1^2)+\frac\pi 2 \alpha(\sigma_2^2 + \sigma_3^2).
\eeqa
Defining the new angle coordinates \cite{GM} such that
\beqa
\sigma_1&=& d\tilde\psi + \cos\tilde\theta d\tilde\phi,\n
\sigma_2&=& -\sin\tilde\psi d\tilde\theta + \sin\theta \cos\tilde\psi d\tilde\phi,\\
\sigma_3&=& \cos\tilde\psi d\tilde\theta + \sin\theta \sin\tilde\psi d\tilde\phi,\nonumber
\eeqa
we see that this metric is regular at $u=0$ ($t=\infty$) if $0\leq\tilde\psi\leq\pi$, 
that is, the points at the angles $\tilde\psi$ and $\tilde\psi+\pi$ must be identified.
This ${\bf Z}_2$ transformation acts on the original angle variables as
\beqa
\theta&\mapsto&\pi-\theta,\n
\phi&\mapsto&\phi+\pi,\\
\psi&\mapsto&-\psi.\nonumber
\eeqa
This means that the space around the Atiyah-Hitchin bolt locus gets ${\bf Z}_2$ orbifolded.
Later this fact places a significant constraint on our construction of the Taub-NUT crystal.

\noindent{\it (ii) The behavior near $t=0$: The negative Taub-NUT}

To investigate the asympotics near $t=0$, one needs to consider the modular 
$S$ transformations of the theta functions \cite{HP}:
\beqa
\vartheta_3\left(0,-\frac1{i\alpha t}\right)&=&\sqrt{\alpha t}~\vartheta_3(0,i\alpha t),\n
\vartheta_4\left(0,-\frac1{i\alpha t}\right)&=&\sqrt{\alpha t}~\vartheta_2(0,i\alpha t),
\label{modularS}\\
\vartheta_2\left(0,-\frac1{i\alpha t}\right)&=&\sqrt{\alpha t}~\vartheta_4(0,i\alpha t).\nonumber
\eeqa
Using (\ref{modularS}), we can similarly obtain
\beqa
w_2(t)&=&
2\frac d{dt}\log\left(\frac1{\sqrt{\alpha t}}~\vartheta_3(0,\frac i{\alpha t})\right)\n
&\stackrel{t\rightarrow 0}{\sim}&
-\frac1t+\cdots,\n
w_3(t)&=&
2\frac d{dt}\log\left(\frac1{\sqrt{\alpha t}}~\vartheta_2(0,\frac i{\alpha t})\right)\n
&\stackrel{t\rightarrow 0}{\sim}&
\frac\pi{2\alpha t^2}-\frac1t+\cdots,\\
w_1(t)&=&
2\frac d{dt}\log\left(\frac1{\sqrt{\alpha t}}~\vartheta_4(0,\frac i{\alpha t})\right)\n
&\stackrel{t\rightarrow 0}{\sim}&
-\frac1t+\cdots,\nonumber
\eeqa 
where the ellipses stand for the terms containing a power of 
$e^{-\frac\pi{\alpha t}}$.
If those contributions are discarded, the $a$, $b$ and $c$ functions reduce to
\beqa
a^2&
\stackrel{t\rightarrow0}{\sim}&
\frac{\pi-2\alpha t}{2\alpha t^2},\n
b^2&
\stackrel{t\rightarrow0}{\sim}&
\frac{\pi-2\alpha t}{2\alpha t^2},\\
c^2&
\stackrel{t\rightarrow0}{\sim}&
\frac {2\alpha}{\pi-2\alpha t}.\nonumber
\eeqa
The metric approaches to
\beqa
ds^2&\stackrel{t\rightarrow0}{\sim}&\frac{\pi-2\alpha t}{2\alpha t^4}dt^2 + \frac {2\alpha}{\pi-2\alpha t} \sigma_3^2
+\frac{\pi-2\alpha t}{2\alpha t^2}(\sigma_1^2 + \sigma_2^2)\n
&=&\left(
1-\frac{\sqrt{\frac{2\alpha}\pi}} r 
\right)\left(d r ^2 +  r ^2(\sigma_1^2 + \sigma_2^2)\right)
+\frac{2\alpha}\pi\left(1-\frac{\sqrt{\frac{2\alpha}\pi}} r \right)^{-1}\sigma_3^2,
\label{negativeT-NUT}
\eeqa
where we have changed variables from $t$ to $ r =(\sqrt{\frac{2\alpha}\pi}t)^{-1}$ in the second line.
(\ref{negativeT-NUT}) is a Taub-NUT metric with  negative NUT charge $-\sqrt{\frac{2\alpha}\pi}$ 
\cite{GM}.

As was noted in \cite{GM}, this approximate metric (\ref{negativeT-NUT})  is singular at 
$ r =\sqrt{\frac{2\alpha}\pi}$; this is an artifact of the absence of the 
exponential terms ignored in the theta functions. On the other hand, the smooth,
self-dual metric (\ref{BianchiIXAH}) rapidly approaches to
the approximate metric (\ref{negativeT-NUT}) as $ r \rightarrow\infty$, and for an observer 
outside a radius of the NUT charge scale, it looks as if there is a Taub-NUT with  negative 
NUT charge, which has an $U(1)$ isometry. 
Later we take T-duality of this 
approximate metric along this isometry.

\section{Taub-NUT crystal}
\subsection{Infinite periodic array of Taub-NUT centers}
In the previous section we have seen that the Atiyah-Hitchin space can be 
approximately identified as a negative-charge Taub-NUT space at large distances 
compared to the NUT charge scale.
Motivated by this observation, in this section we consider the Gibbons-Hawking metric 
for an infinite periodic array of Taub-NUT centers, where some of them in the unit cell 
have  negative NUT charge. 
The idea is that, by periodically identifying such an infinite array, one obtains 
an $S^1$ fibered three-dimensional torus with  (anti-)self-dual metric, 
and by T-duality it is mapped to an NS5-brane system with  compact transverse 
dimensions, where some of the 5-branes have negative tension.

The Gibbons-Hawking metric \cite{GH}
\beqa
ds_{GH}^2&=&V(\vec{x})
d\vec{x}^2 +V^{-1}(\vec{x})(dx_9 + \vec{\omega}\cdot d\vec{x})^2,\label{GH}
%
%
\\
V(\vec{x})&=&\varepsilon+\sum_{k=1}^N \frac{2m}{|\vec{x} -\vec{x}_k |},\label{GHV}\\
\vec\nabla V &=&\vec\nabla \times \vec\omega,\label{GHomega}
\eeqa
is an anti-self-dual metric\footnote{
Whether a metric is self-dual or anti-self-dual is a matter of convention. For instance,
it changes if the ordering of the coordinates is changed from $(\vec{x},x_9)$ to $(x_9, \vec{x})$. 
We have chosen the sign of $\vec{\omega}$ in such a way that $x_9$  is  
naturally identified as $\psi$ (up to a scale, without a sign flip) of the asymptotic 
metric (\ref{negativeT-NUT}). Also, the change of coordinates 
$t\mapsto r =(\sqrt{\frac{2\alpha}\pi}t)^{-1}$ in (\ref{negativeT-NUT}) makes
a self-dual metric  into 
an anti-self-dual one.} in a four-dimensional 
Euclidean space with coordinates $(\vec{x},x_9)$, $\vec{x}=(x,y,z)$. 
We have used the subscript $9$, 
anticipating that this metric will be embedded in string theory. 
$\varepsilon$ and $m$ are constants. If $\varepsilon\neq 0$, 
the metric describes a multi-center Taub-NUT space, while if  $\varepsilon= 0$, 
the  metric is the one for the $A_{N-1}$ singularity \cite{EGH}. If $\varepsilon\neq 0$, 
it can be arranged to $1$ by making a suitable change of coordinates, so 
we can set $\varepsilon= 1$ for the Taub-NUT case without loss of generality.  
Then if $N=1$ and the center $\vec{x}_1$ is taken at the origin $\vec{0}$, 
the potential $V$ takes the form
\beqa
V(\vec{x})=1+\frac{2m}r~~~(r=\sqrt{x^2 + y^2 + z^2}).
\eeqa
Choosing the gauge $\omega_z=0$, $\vec{\omega}$ can be determined 
(for the anti-self-dual case) as
\beqa
\vec{\omega}&=&\left(
-\frac{2myz}{(x^2+y^2)r},\frac{2mxz}{(x^2+y^2)r},0
\right).
\eeqa
Then the metric (\ref{GH}) reads 
\beqa
ds_{GH}^2(N=1)&=&\left(
1+\frac{2m}r
\right)
d\vec{x}^2 +\frac1{
1+\frac{2m}r
}\left(dx_9 + 2m\frac zrd\left(
\tan^{-1}\frac yx
\right)\right)^2.\label{GHN=1}
\eeqa
By a change of coordinates
\beqa
x&=&r \sin\theta \cos\phi,\n
y&=&r \sin\theta \sin\phi,\n
z&=&r \cos\theta,\n
x_9&=&2m\psi,
\eeqa
(\ref{GHN=1}) is reduced to a Taub-NUT metric with  NUT charge $2m$:
\beqa
ds_{GH}^2(N=1)&=&\left(
1+\frac{2m}r
\right)
(dr^2 + r^2(\sigma_1^2+\sigma_2^2)) +4m^2\left(
1+\frac{2m}r
\right)^{-1}
\sigma_3^2.\label{N=1TaubNUT}
\eeqa

Going back to the general $N$ case, 
we impose the following conditions on the configuration of centers:

\noindent
1. For every location of a negative NUT charge $\vec{x}_0$, the configuration 
is invariant under the ${\bf Z}_2$ transformation : 
\beqa
\vec{x}-\vec{x}_0& \mapsto& 
-(\vec{x}-\vec{x}_0)
\label{inversion}
\eeqa
 (called {\it inversion} in the terminology of crystallography).

\noindent
2. The superposition of infinitely many Gibbons-Hawking potentials converges.
 
The first condition is imposed in order for each negative-charge center 
to consistently represent the asymptotic region of the Atiyah-Hitchin geometry. 
To recover correctly the original Atiyah-Hitchin metric 
in the non-compact limit where the negative-charge NUT is isolated from the 
positive ones,
the image of the inversion (\ref{inversion}) needs to be identified, as we saw in 
the previous section.
%

The second condition is obvious;
%
a Gibbons-Hawking potential from a single nut falls like $1/|\vec{x}|$, 
and due to the condition 1 the leading asymptotic behavior of the superposition 
is $1/|\vec{x}|^3$ generically, and therefore  the summation over a 
three-dimensional lattice diverges logarithmically.  The superposition of the potentials 
converges if and only if 
the positive- and negative-charge centers are arranged in a unit cell 
in such a way that the leading order term of the multipole expansion 
vanishes. 
We will present a precise description of the condition in the next section.
Note that the situation is completely analogous 
to the Coulomb static potential in a crystal system, and any actual ionic crystals 
consisting of two oppositely charged ions will satisfy this condition.

In addition, we also assume for simplicity that 

\noindent
3. There are no orbifold fixed points besides the locations of negative charges.

Then a little thought shows that the Gibbons-Hawking potential $V(x)$ takes the form
\beqa
V(\vec{x})&=&
1+\sum_{\rule{0ex}{1.5ex}
n_1 , n_2 , n_3 \in {\bf{Z}}
}
\sum_{\epsilon_1,\epsilon_2,\epsilon_3\in\{0,\frac12\}}
V^{(\epsilon_1,\epsilon_2,\epsilon_3)}({\textstyle 
\vec{x}-\sum_{j=1}^3 n_j  \vec{y}_j}),
\label{VTNC}
\\
V^{(\epsilon_1,\epsilon_2,\epsilon_3)}(\vec{x})&=&
\frac{-4mN^{(\epsilon_1,\epsilon_2,\epsilon_3)}}{\left|\vec{x}-\sum_{j=1}^3 \epsilon_j \vec{y}_j \right|}
+\sum_{k=1}^{N^{(\epsilon_1,\epsilon_2,\epsilon_3)}} 
\left(
\frac{2m}{\left|\vec{x}
-\vec{x}_k^{~(\epsilon_1,\epsilon_2,\epsilon_3)}
-\sum_{j=1}^3 \epsilon_j \vec{y}_j 
\right|}
\right.
\n
&&~~~~~~~~~~~~~~~~~~~~~~~~~~~~~~~~~~~\left.+
\frac{2m}{\left|\vec{x}
+\vec{x}_k^{~(\epsilon_1,\epsilon_2,\epsilon_3)}
-\sum_{j=1}^3 \epsilon_j \vec{y}_j 
\right|}
\right),
\label{Vepsilon}
\eeqa
where ${\vec{y}}_j \in {\bf R}^3$ $(j=1,2,3)$ are the period vectors 
of the array. 

The unit cell of the periodic array is a parallelepiped 
specified by ${\vec{y}}_j$ $(j=1,2,3)$, which contains eight 
($\epsilon_j\in\{0,\frac12\}$ for $j=1,2,3$)
subunits 
(Figure).
\begin{figure}[t]
 \centering
 \includegraphics[scale=0.7]{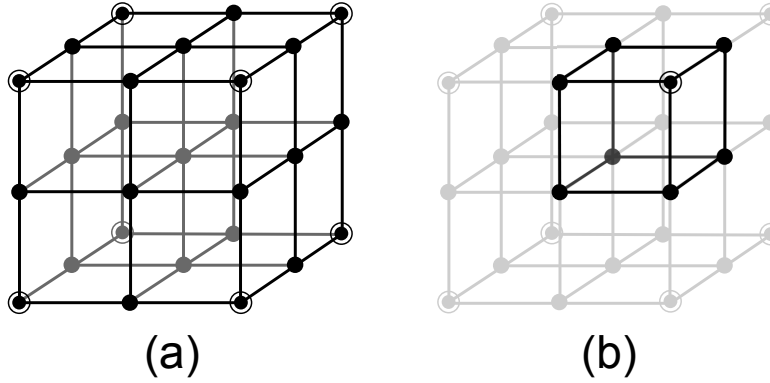}
 \caption{(a) A Unit cell and (b) a subunit of the Taub-NUT crystal. 
 The dots are the positions of negative-charge centers, of which the 
 circled ones are identified periodically. Positive-charge centers, 
 which are not depicted in these figures, must be placed symmetrically 
 with respect to the inversion through any dot. }
 \label{subunit}
\end{figure}
Each subunit  specified by 
$(\epsilon_1,\epsilon_2,\epsilon_3)$ of the array consists of 
one negative-charge center with  NUT charge $-4m N^{(\epsilon_1,\epsilon_2,\epsilon_3)}$ 
and $N^{(\epsilon_1,\epsilon_2,\epsilon_3)}$ pairs of positive-charge centers with 
a common NUT charge $2m$, which are placed symmetrically with respect to 
the position of the negative-charge center. 
The relative positions of the $k$th pair ($k=1,\ldots,,N^{(\epsilon_1,\epsilon_2,\epsilon_3)}$)
are $\pm\vec{x}_k^{~(\epsilon_1,\epsilon_2,\epsilon_3)}$. 
The potential $V^{(\epsilon_1,\epsilon_2,\epsilon_3)}(\vec{x})$ arises from these centers 
belonging to the subunit $(\epsilon_1,\epsilon_2,\epsilon_3)$. 

We now periodically identify points 
as
\beqa
(\vec{x},x_9)&\sim&\left(\vec{x}+\sum_{j=1}^3 m_j   \vec{y}_j,x_9\right)~~~(m_j\in{\bf Z})
\label{translations}
\eeqa
and also take an orbifold
\beqa
(\vec{x},x_9)&\sim&\left(-\vec{x}+2\sum_{j=1}^3( n_j + \epsilon_j)  \vec{y}_j,-x_9\right)
~~~(n_j\in{\bf Z}, ~~\epsilon_j\in{\{0,{\textstyle \frac12}\}}),
\label{inversions}
\eeqa
as we mentioned above.
The metric 
(\ref{GH}) with (\ref{VTNC}),(\ref{Vepsilon}) and (\ref{GHomega}) is invariant under the 
operations (\ref{translations}) and (\ref{inversions}) and the identifications can consistently 
be done.

Note that 
if there were only a single negative-charge center,  (\ref{translations}) 
would identify points in spaces with opposite orientations and the 
quotient would become non-orientable. Also, there would appear 
orbifold fixed points besides their own locations. This is why we have 
introduced eight negative-charge centers in the unit cell.

\subsection{Convergence condition}
The metric (\ref{GH}) with (\ref{VTNC}),(\ref{Vepsilon}) and (\ref{GHomega}) 
 is an infinite, periodic generalization of the Gibbons-Hawking 
metric. In this section we consider in what circumstances the potential 
$V$ converges. 
If the positive and negative centers are in generic positions, 
the potential (\ref{Vepsilon}) falls like $1/|\vec{x}|^3$ since it can be written as a sum 
of second order differences, and therefore the summation over a three-dimensional 
lattice diverges logarithmically. However, the exception to this is where 
the leading order term of the multipole expansion of
\beqa
\sum_{\epsilon_1,\epsilon_2,\epsilon_3\in\{0,\frac12\}}
V^{(\epsilon_1,\epsilon_2,\epsilon_3)}(\vec{x})
\label{sumVepsilon}
\eeqa
vanishes, that is,  
the positive- and negative-charge centers are arranged in a unit cell 
in such a way that the leading difference operator of the multipole expansion 
is proportional to the discrete Laplace operator. 

\begin{figure}[h]
 \centering
 \includegraphics[scale=0.7]{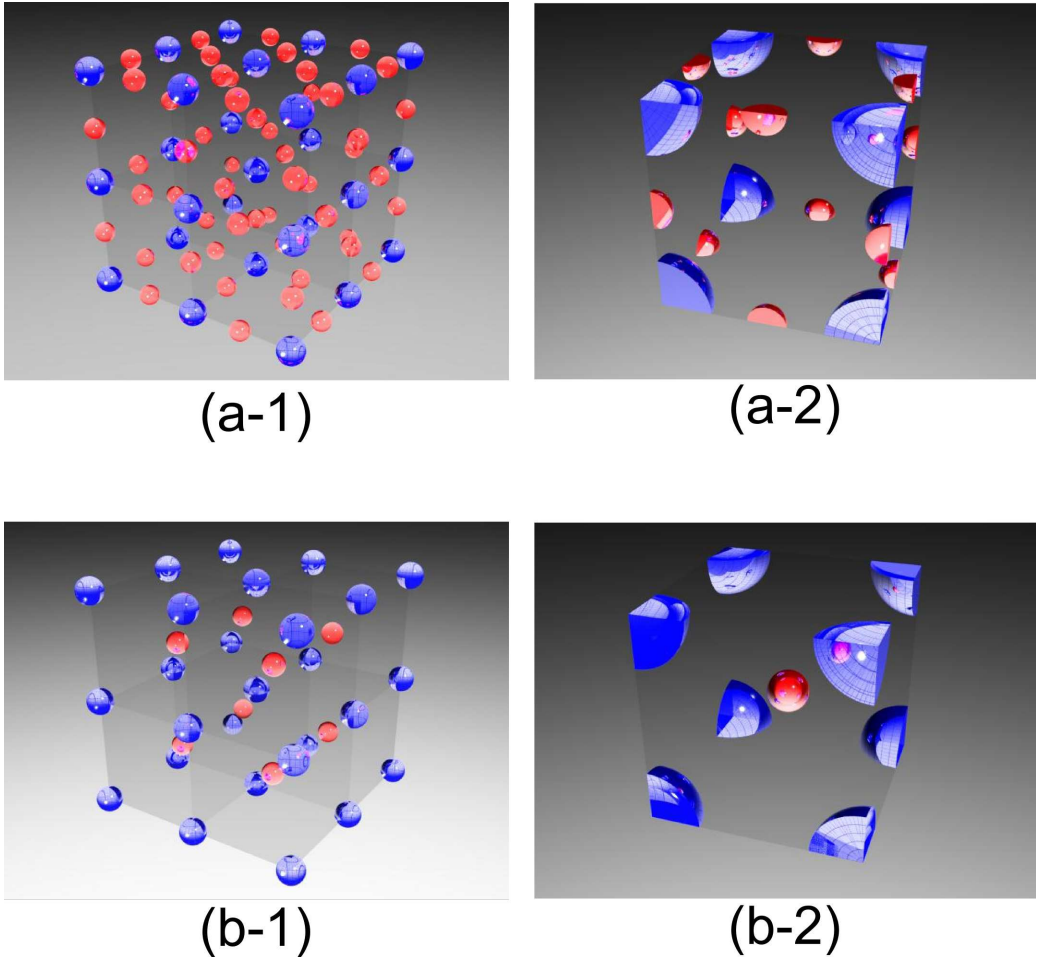}
 \caption{Examples of convergent configurations. 
  The unit cells of  (a-1) the $ReO_3$-type and 
  (b-1) the $CsCl$-type crystals are depicted, and 
 (a-2), (b-2) are their sub-units, respectively. The blue spheres show the 
  positions of the negative-charge (``Atiyah-Hitchin") centers, whereas the 
  red ones are the positive-charge (``Taub-NUT") centers. }
 \label{figReO3CsCl}
\end{figure}

The simplest example of a convergent potential is the case where 
the lattice is a cubic lattice
\beqa
\vec{y}_j=2a \vec{n}_j~~~(j=1,2,3),
\eeqa
\beqa
\vec{n}_1\equiv(1,0,0),~\vec{n}_2\equiv(0,1,0),~\vec{n}_3\equiv(0,0,1),
\eeqa
and
all $V^{(\epsilon_1,\epsilon_2,\epsilon_3)}(\vec{x})$'s are obtained as
translations of an identical potential: 
\beqa
V^{(\epsilon_1,\epsilon_2,\epsilon_3)}(\vec{x}) &=&
V_{ReO_3}(\vec{x}-{\textstyle \sum_{j=1}^3 \epsilon_j \vec{y}_j},\delta),\\
V_{ReO_3}(\vec{x},\delta)&\equiv&2m\left(
-\frac{6}{\left|\vec{x}\right|}
+\sum_{j=1}^3
\left(
\frac1{|\vec{x}-\delta a \vec{n}_j|}
+\frac1{|\vec{x}+\delta a \vec{n}_j|}
\right)
\right),
\label{VReO3}
\eeqa
where $a$ is half the period of the array and $0< \delta \leq 1$ is a parameter
(Figure \ref{figReO3CsCl}  (a-1), (a-2)). 
\footnote{The actual crystal of rhenium trioxide $(ReO_3)$
realizes  this configuration for $\delta=1$. 
Strictly speaking, if the Taub-NUT positions change from $\delta\neq 1$ to 
$\delta=1$ and two Taub-NUT centers come on top of each other, then 
the periodicity of the $x_9$ direction must be changed so as to avoid 
the NUT singularity. Here, we are interested in the convergence of the three-dimensional 
potential and ignore this complication.
}
%

Indeed, we can rewrite (\ref{VReO3}) as
\beqa
&&2m\left(-\frac{6}{\left|\vec{x}\right|}
+\sum_{j=1}^3
\left(
\frac1{|\vec{x}-\delta a \vec{n}_j|}
+\frac1{|\vec{x}+\delta a \vec{n}_j|}
\right)
\right)\n
&=&
2m
\sum_{j=1}^3
\left(\left(
\frac1{|\vec{x}+\delta a \vec{n}_j|}-\frac{1}{\left|\vec{x}\right|}
\right)-
\left(\frac{1}{\left|\vec{x}\right|}
-\frac1{|\vec{x}-\delta a \vec{n}_j|}
\right)
\right)\n
&=&
2m\delta^2 a^2
\sum_{j=1}^3
(\vec{n}_j\cdot
\vec{\nabla})^2 
\frac1{|\vec{x}|}
+O(a^4)
\n
&=&
2m\delta^2 a^2
\Delta\frac1{|\vec{x}|}
+O(a^4),
\eeqa
where $\Delta$ is the flat space Laplacian. The first term vanishes, and 
on dimensional grounds the rest falls like $|\vec{x}|^{-5}$ at least.
Therefore the summation over the lattice converges.

Another simple example of a convergent potential is the case in which 
the lattice is again cubic and a subunit has four (pairs of) Taub-NUT's per single 
Atiyah-Hitchin (Figure \ref{figReO3CsCl}  (b-1), (b-2)): 
\beqa
V^{(\epsilon_1,\epsilon_2,\epsilon_3)}(\vec{x}) &=&
V_{CsCl}(\vec{x}-{\textstyle \sum_{j=1}^3 \epsilon_j \vec{y}_j},\delta),\\
V_{CsCl}(\vec{x},\delta)&\equiv&2m\left(
-\frac{8}{\left|\vec{x}\right|}
+\sum_{\xi_1,\xi_2,\xi_3=\pm1}
\frac1{\left|\vec{x}-\delta  a \sum_{j=1}^3\xi_j\vec{n}_j\right|}
\right).
\label{VCsCl}
\eeqa

More generally, if one requires that for each subunit $(\epsilon_1,\epsilon_2,\epsilon_3)$ 
the leading-order term of $V^{(\epsilon_1,\epsilon_2,\epsilon_3)}(\vec{x})$  
in the large-$|\vec{x}|$ expansion vanishes,  the condition for the relative positions 
of the Taub-NUT centers is that the $3\times N^{(\epsilon_1,\epsilon_2,\epsilon_3)}$-matrix
\beqa
X^{(\epsilon_1,\epsilon_2,\epsilon_3)}&\equiv&\left(
\vec{x}_1^{(\epsilon_1,\epsilon_2,\epsilon_3)},
\vec{x}_2^{(\epsilon_1,\epsilon_2,\epsilon_3)},
\ldots,
\vec{x}_{N^{(\epsilon_1,\epsilon_2,\epsilon_3)}}^{(\epsilon_1,\epsilon_2,\epsilon_3)}
\right)
\eeqa
should satisfy
\beqa
X^{(\epsilon_1,\epsilon_2,\epsilon_3)}{X^{(\epsilon_1,\epsilon_2,\epsilon_3)}}^T
&=&kI_3
\label{XX=kI}
\eeqa
for some constant $k$, where $I_3$ is the $3\times 3$ identity matrix.
It is easy to see that the two examples above fulfill this condition.

Note that (\ref{XX=kI}) is a sufficient condition for $V(\vec{x})$ to converge;
it is enough if the {\em sum} (\ref{sumVepsilon}) over the unit has the asymptotic fall-off is 
not slower than $|\vec{x}|^{-3}$, and each $V^{(\epsilon_1,\epsilon_2,\epsilon_3)}(\vec{x})$ 
need not subunit-wise. Of course, in such a general case the crystal has less discrete 
symmetries, however. 

\subsection{Taub-NUT crystal as an approximation of $K_3$}\label{K3}
The negative-charge metric (\ref{negativeT-NUT}) is singular, but adding exponential 
correction terms the singularity disappears and the metric becomes completely smooth 
to be the Atiyah-Hitchin metric.  These exponential corrections are regarded as coming 
from certain instantons \cite{SW,Doreyetal,HP}. In the language of the corresponding 
gauge theory, this is a strong-coupling effect of the non-abelian ($=SU(2)$) gauge 
theory at low energies, where the location of the singularity surface corresponds to 
$\Lambda_{QCD}$. It is reasonable to expect that there will also be such corrections 
even after the three-dimensional lattice is compactified to be a torus. If this is true, and 
if these corrections can smooth out the singularity as they do in the non-compact case, 
then the resulting space must be a $K_3$ surface, as it is the only nontrivial smooth 
compact hyper-K\"{a}hler four-manifold. 
The compactified periodic array can then be regarded as an approximation of 
a $K_3$ surface  (which is only valid except near the locations of the negative-charge 
centers). 
However, this can only happen if 
\beqa
\sum_{\epsilon_1,\epsilon_2,\epsilon_3\in\{0,\frac12\}}
N^{(\epsilon_1,\epsilon_2,\epsilon_3)}&=&16
\eeqa
in a unit cell since the Euler number of $K_3$ is 24. This means that (apparently, 
including their ${\bf Z}_2$ images) there are 32 positive-charge Taub-NUT centers 
in a unit cell before the ${\bf Z}_2$ identification. Note that although the number of 
negative-charge centers is apparently eight, it must be halved when counting 
the number of bolts due to the ${\bf Z}_2$ identification. (This is the same as the Euler-number
counting of the $K_3$ orbifold \cite{Walton}, 
where the number of excised points is not equal to the number of the 
fixed points itself, but it must be divided by the order of the orbifold.)
Therefore, since  a single nut adds one to 
the Euler number while a bolt adds two to it, the total Euler number amounts to 
\beqa
1\times 16~+~2\times \frac82 ~=~ 24.
\eeqa
In this case, the compactified and orbifolded periodic array is related by 
a chain of dualities to the $K_3$ orientifold, where the sixteen Taub-NUT's here 
correspond to the same number of D9-branes.

\subsection{Classification of the crystal structure}
\begin{figure}[b]
 \centering
 \includegraphics[scale=0.7]{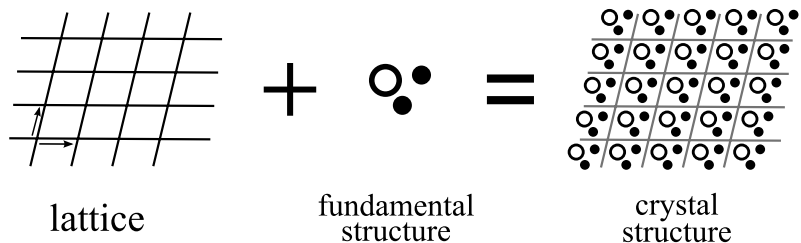}
 \caption{The crystal structure.}
 \label{figA}
\end{figure}

\begin{figure}[t]
 \centering
 \includegraphics[scale=1]{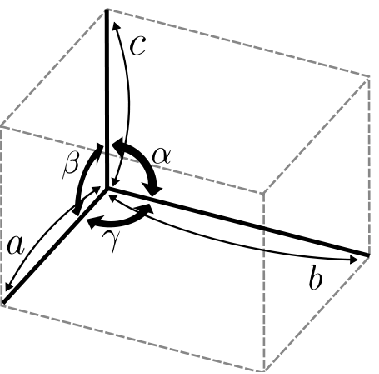}
 \caption{The definitions of the edge lengths and angles in Table~\ref{figC}.}
 \label{figB}
\end{figure}

Crystals are known to be classified by their symmetries.
The crystal structure is characterized by the following two aspects: One
is the {\em lattice} that determines its translational 
symmetries, and the other is the {\em fundamental structure} whose 
periodical repetitions generated by the lattice yields the 
whole crystal (Figure \ref{figA}).
%
Three-dimensional 
crystals fall into seven classes, the {\em crystal systems},  
depending the shape of the parallelepiped of the unit cell 
that the crystal lattice specifies. Four of the seven crystal systems,
the monoclinic, orthorhombic, tetragonal and cubic systems, are 
further divided into several sub-classes, as shown in Table~\ref{figC}.
The definitions of the edge lengths and angles of the parallelepiped are
shown in Figure~\ref{figB}.

%
\newlength{\myheight}
\setlength{\myheight}{2cm}
\newlength{\myheighta}
\setlength{\myheighta}{1cm}
\newcommand{\bhline}[1]{\noalign{\hrule height #1}}
\newcommand{\bvline}[1]{\vrule width #1}
\begin{center}
 \begin{table}[hbtp]
 \begin{tabular}{@{\bvline{1pt}}c|c|c|c|c|c@{\bvline{1pt}}}
 \bhline{1pt}
  &\raisebox{-1mm}{\{length, angles\}} &
 \multicolumn{4}{|c@{\bvline{1pt}}}{Bravais lattice}\\
 \cline{3-6}
 \raisebox{4mm}{\, crystal system} & 
                      \raisebox{1mm}{\{symmetry element\}}
                                                              & P & C & I & F \\
 \bhline{1pt}
   
 \parbox[c][\myheight][c]{0cm}{}
 triclinic & 
        $
        \begin{array}{c}
                  \{ a \ne b \ne c, \alpha \ne \beta \ne \gamma \} \\
                  \{ \mbox{no symmetry element} \}
         \end{array}$
      &  \raisebox{-5mm}{\includegraphics[scale=0.7]{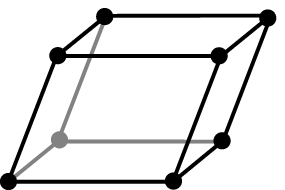}} & - & - & - \\
 \hline
 \parbox[c][\myheight][c]{0cm}{}
 monoclinic & 
                          $\begin{array}{c}
                               \left\{\begin{array}{c}       
                                 a \ne b \ne c \\
                                 \alpha = \gamma = \pi/2 ,\beta \geq \pi/2
                                 \end{array} \right\}
                                 \\
                                \{\mbox{ 1-diad}\}
                                \end{array} $ &  \raisebox{-7.5mm}
                                                 {\includegraphics[scale=0.7]{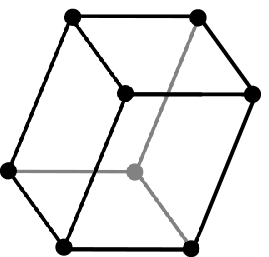}} & 
                                                 \raisebox{-7.5mm}
                                                 {\includegraphics[scale=0.7]{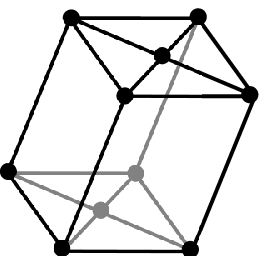}} &
                                                                     -                       &
                                                                     -                        \\ 
 \hline
 \parbox[c][\myheight][c]{0cm}{}
 orthorhombic &
                                        $       \begin{array}{c}
                                                 \left\{\begin{array}{c}
                                                 a \ne b \ne c\\
                                                \alpha = \beta = \gamma = \pi/2
                                                        \end{array}\right\}\\
                                                \{\mbox{mutually orthogonal 3-diad}\} 
                                                \end{array} $ & \raisebox{-7.5mm}{\includegraphics[scale=0.7]{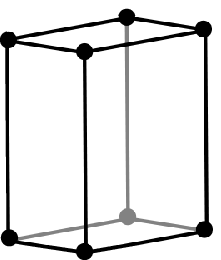}} &
                                                                \raisebox{-7.5mm}{\includegraphics[scale=0.7]{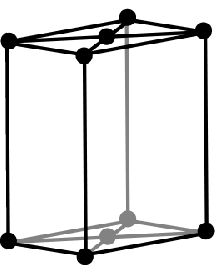}} &
                                                                \raisebox{-7.5mm}{\includegraphics[scale=0.7]{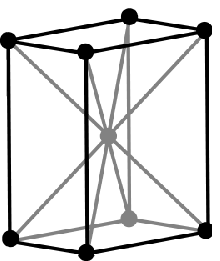}} &
                                                                \raisebox{-7.5mm}{\includegraphics[scale=0.7]{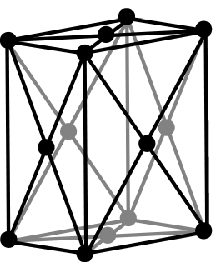}} \,  \\
 \hline
 \parbox[c][\myheight][c]{0cm}{}
  hexagonal &                             $    \begin{array}{c}
                                                \left\{\begin{array}{c}
                                                 a = b \ne c\\
                                                \gamma = 2\pi/3 
                                                 \end{array}\right\}\\
                                                \{\mbox{1-hexad}\} 
                                                \end{array}  $ & \raisebox{-6mm}{\includegraphics[scale=0.7]{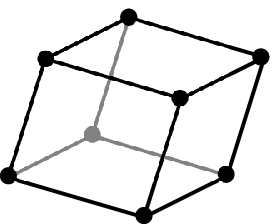}} & - & - & - \\
 \hline
 \parbox[c][\myheight][c]{0cm}{}
 trigonal &                                    $\begin{array}{c}
                                                 \left\{\begin{array}{c}
                                                 a = b = c\\
                                                \gamma \ne \pi/3 , \pi/2 ,
                                                \cos^{-1}(-{1\over3}) 
                                                        \end{array}\right\}\\
                                                \{\mbox{1-triad}\}
                                                \end{array} $ & \raisebox{-5.5mm}{\includegraphics[scale=0.7]{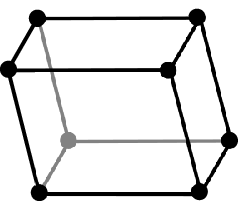}} & - & - & - \\
 \hline
 \parbox[c][\myheight][c]{0cm}{}
 tetragonal &                                  $\begin{array}{c}
                                                \left\{\begin{array}{c}
                                                 a = b \ne c\\
                                                \alpha = \beta = \gamma = \pi/2
                                                \end{array}\right\}\\
                                                \{\mbox{1-tetrad}\} 
                                                \end{array} $ & \raisebox{-8.5mm}{\includegraphics[scale=0.7]{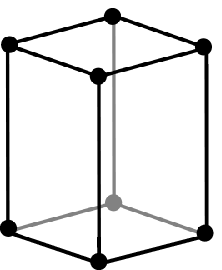}} &
                                                                                       -                    &
                                                                \raisebox{-8.5mm}{\includegraphics[scale=0.7]{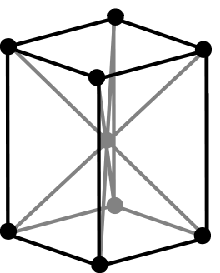}} &
                                                                                       -                     \\
 \hline
 \parbox[c][\myheight][c]{0cm}{}
 cubic &                                       $\begin{array}{c}
                                                 \left\{\begin{array}{c}
                                                 a = b = c\\
                                                \alpha = \beta = \gamma = \pi/2
                                                 \end{array}\right\}\\
                                                \{\mbox{4-triad}\} 
                                                \end{array} $ & \raisebox{-6mm}{\includegraphics[scale=0.7]{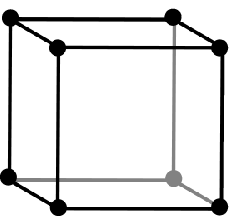}} &
                                                                                    -                  &
                                                                \raisebox{-6mm}{\includegraphics[scale=0.7]{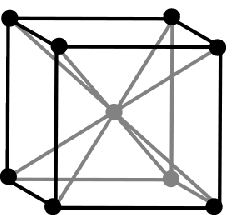}} &
                                                                \raisebox{-6mm}{\includegraphics[scale=0.7]{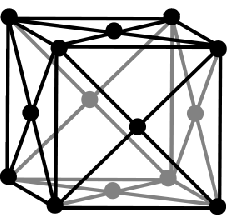}} \, \\
 
 \bhline{1pt}
 \end{tabular}
 \caption{The Bravais lattices.}
  \label{figC}
   \end{table}
\end{center}
%

%

Two crystal lattices belonging to the same crystal system but different 
sub-classes (that is, two ones in the same row in Table \ref{figC}) 
possess the same point group symmetries 
(that is, 
the group of symmetry operations that leave at least one lattice point 
fixed)
but different 
translational symmetries. These, in all, fourteen crystal lattices are 
known as the {\em Bravais lattices} in crystallography.
The classification into the fourteen Bravais lattices is further refined 
into 230 crystal types according to their discrete symmetries called 
{\em space groups}, which are generated by the point group and the 
group of translational symmetries.  

One of the important aspects of such discrete symmetries of a crystal 
is that they might be used to explain an origin of discrete flavor symmetries 
\cite{discrete_flavor_symmetries} in nature (if any), though in the present 
case in the compactification to six dimensions. Indeed, it has been known 
for a long time that there are localized \cite{CHS} 30 hypermultiplets 
on each symmetric 5-brane (obtained by T-duality in the next section), 
whose existence is ensured by anomaly cancellation \cite{anomaly,KM2},
and these hypermultiplets are to be observed as flavors in the warped 
compactification. 
It would be interesting to generalize the argument to a similar periodic array of 
intersecting 5-branes.

\begin{figure}[h]
 \centering
 \includegraphics[scale=0.8]{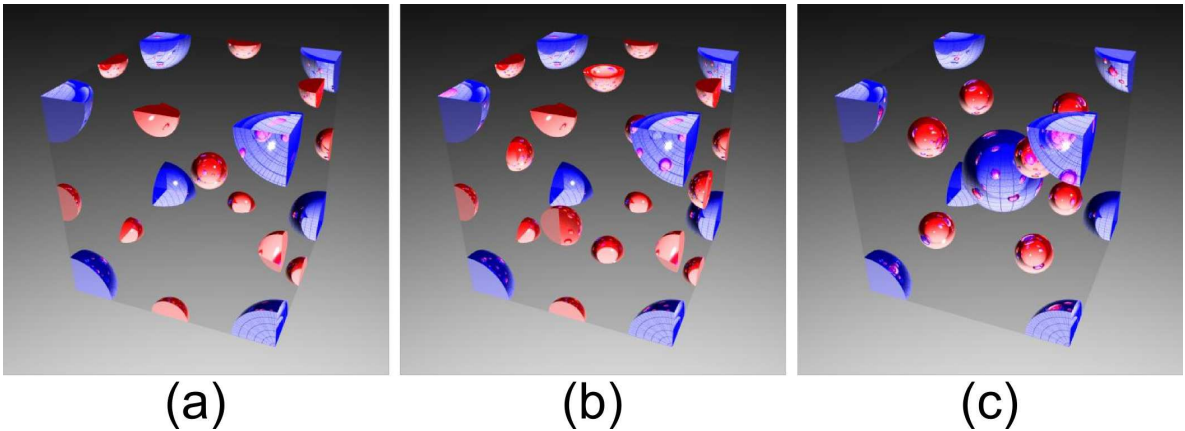}
 \caption{Cubic lattices with triads that have 16 positive-charge 
Taub-NUT's. 
 (a) and (b) are of the P type and their subunits are shown there.
 (c) is of the I type and hence a region larger than a subunit is 
 depicted so that its symmetry may be manifest.}
 \label{cubic_triad}
\end{figure}

Taub-NUT crystals with a highly symmetric structure are very limited.
For instance, let us consider cubic lattices with ``triads" (that is, 
symmetry axes of rotations by $120$ degrees) that have 16 positive-charge 
Taub-NUT's
(Euler number $=24$) 
as in section \ref{K3}. 
Then it turns out that the possible type is either of the three shown in Figure 
\ref{cubic_triad}.

%
%

%

\section{T-duality to NS5-brane system}
\subsection{$O(1,1)$ dual and embedding to heterotic string theory}
It is also a well-known fact that  the NS5-brane \cite{CHS,SJRey} 
and the Taub-NUT space 
are related by T-duality \cite{GHM,ACL}, which we briefly review here. 
The NS-NS sector Lagrangian is given in the string frame as
\beqa
2\kappa^2{\cal L}_{NSNS}&=&Ee^{-2\Phi}\left(
R+4\partial_M \Phi \partial^M \Phi
-\frac1{12}H_{MNP}H^{MNP}
\right),
\label{stringframe}
\eeqa
where $E$ is the determinant of the vielbein $E^{~A}_{M}$, related 
with the string-frame metric $G_{MN}$ $(M,N=0,\ldots,9)$ as
$G_{MN}=E^{~A}_{M}E^{~B}_{N}\eta_{AB}$.
Using the Lorentz invariance, one can cast the vielbein into the form
\beqa
E^{~A}_{M}&=&\left(
\begin{array}{cc}
E^{~\alpha}_{\mu}& A_\mu \widehat{E}\\
&\\
0&\widehat{E}
\end{array}
\right),
\eeqa
where $\mu$ ($\alpha$) is a nine-dimensional curved (flat) index.
If the metric and all other fields do not depend on $x^9$,
then decomposing the vielbein as
the Lagrangian can be reduced to an $O(1,1)$ sigma model coupled to 
gravity, an $B$ field, two abelian
vector fields and a dilaton, which is 
invariant under the following transformation \cite{MSSen}:
\begin{alignat}{3}
&{E^{\rm new}}^{~\alpha}_{\mu} & 
&=& \quad
&E^{~\alpha}_{\mu},\n
&{A^{\rm new}}_\mu& 
&=& \quad 
& B_{\mu9},\n
&{\widehat{E}}^{\rm new}&
&=& \quad
&\widehat{E}^{-1},\label{O(1,1)}\\
&{B^{\rm new}}_{\mu\nu}&
&=& \quad
&B_{\mu\nu}-2 A_{[\nu}B_{\mu]9},\n
&{B^{\rm new}}_{\mu9}&&=&\quad
&A_\mu,\n
&{\Phi}^{\rm new}&&=&
\quad&\Phi-\log\widehat{E}.\nonumber
\end{alignat}
Therefore, starting from a solution of the equations of motion,
one can obtain another new solution by applying this transformation to 
the old one.

We apply the T-duality transformation to 
the Gibbons-Hawking multi-center metric \cite{GH}.
If the Gibbons-Hawking metric (\ref{GH}) is embedded in eleven dimensions and 
the $x_9$ direction is compactified on $S^1$, the metric describes $N$ 
D6-branes in type IIA theory; this is a string theory analogue of the GPS 
monopole. Alternatively, we embed (\ref{GH})  as a four-dimensional 
metric in ten dimensions so that the vielbein becomes 
\beqa
E_M^{~A}&=&\left(
\begin{array}{ccc}
{\bf 1}_6&0&0\\
 0&V^{\frac12}{\bf 1}_3&V^{-\frac12}\vec{\omega}\\
 0&0&V^{-\frac12}
\end{array}
\right),
\eeqa
while other NS-NS fields are set to zero.
Then the $9+1$ decomposition reads
%
\begin{alignat}{3}
&{E}^{~\alpha}_{\mu} & 
&=& \quad
&\left(
\begin{array}{cc}
{\bf 1}_6&0\\
 0&V^{\frac12}{\bf 1}_3
\end{array}
\right),\n
&A_{\mu}& 
&=& \quad 
& \left\{\begin{array}{ll}
0&(\mu=0,\ldots,5),\\
\omega_x&(\mu=6),\\
\omega_y&(\mu=7),\\
\omega_z&(\mu=8),
\end{array}\right.\n
&{\widehat{E}}&
&=& \quad
&V^{-\frac12},\\
&{B}_{\mu\nu}&
&=& \quad
&0,\n
&{B}_{\mu9}&&=&\quad
&0,\n
&{\Phi}&&=&
\quad&0.\nonumber
\end{alignat}
Using the transformation rules (\ref{O(1,1)}), we obtain the T-dual 
configurations as 
\begin{alignat}{3}
&{E^{\rm new}}^{~\alpha}_{\mu} & 
&=& \quad
&\left(
\begin{array}{cc}
{\bf 1}_6&0\\
 0&V^{\frac12}{\bf 1}_3
\end{array}
\right),\n
&{A^{\rm new}}_\mu& 
&=& \quad 
& 0,\n
&{\widehat{E}}^{\rm new}&
&=& \quad
&V^{\frac12},\label{newfields}\\
&{B^{\rm new}}_{\mu\nu}&
&=& \quad
&0,\n
&{B^{\rm new}}_{\mu9}&&=&\quad
& \left\{\begin{array}{ll}
0&(\mu=0,\ldots,5),\\
\omega_x&(\mu=6),\\
\omega_y&(\mu=7),\\
\omega_z&(\mu=8),
\end{array}\right.\n
&{\Phi}^{\rm new}&&=&
\quad&\frac12\log V,\nonumber
\end{alignat}
or
\beqa
ds_{\rm new}^2&=&
\eta_{ij} dx^i dx^j 
+
V(\vec{x}) \delta_{\mu\nu} dx^\mu dx^\nu,\label{smearedNS5V}
\\
H^{\rm new}_{\mu\nu 9}&=&\epsilon^\rho_{~\mu\nu 9 }\partial_\rho V(\vec{x}),\label{smearedNS5H}
\\
e^{2\Phi^{\rm new}}&=&V(\vec{x}),\label{smearedNS5Phi}
\eeqa
where $\epsilon^\rho_{~\mu\nu 9 }$ is normalized to take values $\pm 1$.
(\ref{smearedNS5V})-(\ref{smearedNS5Phi})
describe $N$ smeared NS5-branes.

It is now obvious that, to obtain the T-dual of the infinite periodic array that 
we described in section 3, 
we have only to replace $V$ with (\ref{VTNC}), (\ref{Vepsilon}) in 
(\ref{smearedNS5V})-(\ref{smearedNS5Phi}), as the T-dual transformation does not 
depend on the explicit form of $V(\vec{x})$.
Thus we obtain a three-dimensional periodic array of, in this case, parallel NS5-branes 
with one ($x^9$) dimension smeared and compactified on a circle.  
The ${\bf Z}_2$ projection (\ref{inversions}) must also be imposed. 

It can also be easily verified that $\Omega^{\rm new}_{\mu\alpha\beta}\pm 
H^{\rm new}_{\mu\alpha\beta}$, 
where $\Omega^{\rm new}_{\mu\alpha\beta}$ is the spin connection derived from 
${E^{\rm new}}^{~\alpha}_{\mu}$, belong to different $SU(2)$ subalgebras  
of the transverse space holonomy group $SO(4)$. Therefore, the 5-brane system 
has a hyper-K\"{a}hler with torsion geometry, and hence preserves one half the 
supersymmetries. We can then construct in the standard fashion a supersymmetric 
heterotic background by identifying some part of the gauge connection to be 
equal to the generalized spin connection:
\beqa
\Omega^{\rm new}_{\mu\alpha\beta}\pm H^{\rm new}_{\mu\alpha\beta}=(A_\mu)_{\alpha\beta},
\eeqa 
where $\alpha,\beta$ on the right hand side are the gauge indices specifying 
some $SU(2)$ subalgebra of $E_8\times E_8$ or $SO(32)$.

\subsection{T-dual of negative-charge Taub-NUT as heterotic orientifold}
The ${\bf Z}_2$ identification through every location of the ``Atiyah-Hitchin" 
brane shows that they are orientifold-like objects in heterotic string theory. 
Another property which supports the identification is their tensions: 
In the string frame, they have negative ADM energy per unit area as can be 
easily verified. 
Of course, this is made possible because they have singularities.

In order to construct a 5-brane system with compact 
transverse dimensions, we first considered an infinite periodic array 
of Taub-NUT centers. To make the Gibbons-Hawking potential converge, 
we were naturally led to introduce negative charge centers as a part of it. 
Then by 
T-duality, they turned into branes with negative tension. That is, they need to 
appear, in this construction, for the transverse space to be compactified.
This may be viewed as in accordance with the statement of \cite{GKP}.      

Also, 
in \cite{KM2,KM}, a brane compactification of 
the $E_8\times E_8$ heterotic string theory was considered using smeared 
intersecting 5-branes and the spectrum of fermionic zeromodes was computed.
There, some of the intersecting branes were required to have negative tension 
for a convergent harmonic function of the metric. However, without a microscopic 
definition of orientifolds in heterotic string, their origin was obscure. 
The T-dual of (the asymptotic form of) the Atiyah-Hitchin space thus may
provide an understanding of such negative tension branes 
in heterotic string theory (despite the singularity; see next section).

\subsection{Discussion: Vortex corrections to the singularity}
So far we have considered a system of brane compactification  
of string theory 
in the framework of supergravity. In any case, it is not useful 
when the string coupling is not small, or when the fields vary rapidly compared to 
the string scale. In this section we discuss the possibility of probing the geometry near the 
5-branes by using the sigma-model approach.  
Of course, heterotic string 
has no D-brane probes, but one may use them first in type II theories as a systematic  
means of obtaining a hyper-K\"{a}hler geometry (with torsion), and then it may be 
converted into a heterotic background by the standard embedding.

The three-dimensional $N=4$, $SU(2)$ supersymmetric gauge theory obtained 
by a dimensional reduction of the four-dimensional $N=2$ gauge theory without 
hypermultiplets has the Atiyah-Hitchin space as its moduli space \cite{SW,Doreyetal}.
In three dimensions, the $U(1)$ gauge field is dualized to a scalar, so that there are 
four real scalars whose expectation values parameterize the 
Coulomb branch moduli space, which must be some hyper-K\"{a}hler manifold 
\cite{A-GF}. To one-loop, the metric is nothing but the negative-charge Taub-NUT 
that we have used in the Gibbons-Hawking metric. By taking into account the 
corrections due to monopoles, it was shown that the smooth Atiyah-Hitchin space 
was singled out from other hyper-K\"{a}hler manifolds \cite{SW,Doreyetal}.
Originally, the Atiyah-Hitchin space was discovered \cite{AH} as the moduli space of 
slowly-moving two monopoles in a four-dimensional gauge theory \cite{Manton}. 
The appearance of the same moduli space of different gauge theories was explained 
\cite{HW} by considering a D3-NS5-brane system, where the two gauge theories 
arise as low-energy theories on different branes. 

Now suppose that  we consider a further dimensional reduction to two dimensions. 
This is a two-dimensional $(4,4)$ simga model, consisting of a hypermultiplet and a 
twisted hypermultiplet, both in the adjoint of $SU(2)$. In two dimensions the vector 
field have no propagating degrees of freedom,  and there is no need to dualize 
anything. The degree of freedom that the three-dimensional dual scalar had is now 
carried by one of the scalar components of the twisted hypermultiplet. 
This is essentially the Buscher's T-duality, and the sigma model is thus expected to 
describe the T-dual geometry  
of (in the $SU(2)$ case) the Atiyah-Hitchin space. 

This set up was already considered long time ago by Diaconescu and Seiberg \cite{DS}.
Indeed, in the $SU(2)$ case, they obtained a one-loop (exact) sigma model metric 
precisely the same as the singular negative-tension brane that we obtained as
the T-dual of the negative-charge Taub-NUT. (See also \cite{Smilga}.)  The relative 
minus sign in the harmonic function is a reflection of the non-abelian nature of 
the gauge theory, and in the $U(1)$ case the relative sign is positive, and that 
corresponds to the ordinary Taub-NUT space.  However, unlike in three dimensions 
where monopoles can exist, 
there are, at least in the naive sense, no further non-perturbative contribution 
from vortices 
in two dimensions on this Coulomb branch;  there is no vortices 
in the reduced two-dimensional theory of the pure $SU(2)$ gauge theory broken to $U(1)$ 
since $\pi_1(SU(2)/U(1))$ is trivial.  Still, we note that it has also been known for some 
time that, in a sense described in \cite{MNS}, the vortex moduli space of the Hitchin system is 
well-defined and its volume can be computed \cite{MNS}. 

Realizing the hyper-K\"{a}hler geometry with torsion on the Higgs branch is also 
interesting. In \cite{Tong}, the smeared 5-brane metric was realized as the Higgs branch 
moduli space of a $U(1)$ $(4,4)$ sigma model, and it was argued that the world sheet 
instantons ($=$ vortices) correct the metric into the Poisson re-summed form of 
the localized 5-brane metric compactified on a circle. It would be interesting if one could 
generalize this to the negative-tension/Atiyah-Hitchin case.  
The hyper-K\"{a}hler quotient 
construction for the singular two-monopole moduli space ($=$ the Atiyah-Hitchin 
space without instanton corrections) is already known \cite{GR}. What puzzles us about 
the result is that, in that case, one needs to start from a flat hyper-K\"{a}hler space 
with an {\em indefinite} metric, which would mean that some of the scalars 
of the sigma model would have to have a wrong-sign kinetic term. The realization on
the Higgs branch requires further study.


\section{Summary}
In this paper we have taken a first step toward realizing warped compactifications 
using NS5-branes in heterotic string theory. We have first considered the Gibbons-Hawking 
metric for an infinite periodic array of Taub-NUT's, and have taken T-duality 
to obtain a parallel 5-brane system with a compact transverse space having the 
$SU(2)$ structure, which allowed us to obtain a supersymmetric heterotic 
5-brane system via the standard embedding. In order for the Gibbons-Hawking potential 
to converge, there have been included Taub-NUT centers having a negative NUT 
charge, which have been identified as the asymptotic forms of the Atiyah-Hitchin metric. 
This identification has forced us to make a ${\bf Z}_2$ projection 
through the positions 
of the negative Taub-NUT centers. We have described the convergence condition 
for the locations of the centers. We have shown that in order for the periodic array to be 
an approximation of $K_3$, the number of the positive charge Taub-NUT's must be 16 in 
a unit cell. We have seen some parallels between the periodic array of the Taub-NUT 
centers and ionic crystals, and structures of the ``Taub-NUT crystal" can be 
similarly classified in terms of the Bravais lattices, of which we have given a brief review.

Since we have used the asymptotic form of the Atiyah-Hitchin metric as a part of 
the construction, the metric 
(\ref{GH}) with (\ref{VTNC}),(\ref{Vepsilon}) and (\ref{GHomega})
 have singularities 
near the locations of the negative charges \cite{GM,Sen}, and it only makes sense 
at distances larger than the charge scale from those charges. 
Despite the singularities, the compactified/orbifolded
infinite periodic array 
has significance in that (i) it clarifies the origin of negative tension branes 
in heterotic string theory, and (ii) it basically gives the leading-order behavior 
of such objects whose corrections from nonperturbative effects might be 
systematically analyzed by the sigma model approach as we discussed in 
the text.

\section*{Appendix ~~~ The Darboux-Halphen system}
\setcounter{equation}{0}
\renewcommand\theequation{A.\arabic{equation}}

\subsection*{A.1~ The Bianchi IX self-dual metrics}

The Bianichi IX metric:
\beqa
ds^2= a^2 b^2 c^2 dt^2 + a^2 \sigma_1^2 + b^2 \sigma_2^2 + c^2 \sigma_3^2.
\label{BianchiIXA}
\eeqa
$a$, $b$ and $c$ are functions of $t$ only.  $\sigma_i$'s $(i=1,2,3)$ are 
the Maurer-Cartan 1-forms of $SU(2)$ given by
\beqa
\sigma_1&=& -\sin\psi d\theta + \sin\theta \cos\psi d\phi,\n
\sigma_2&=& \cos\psi d\theta + \sin\theta \sin\psi d\phi,\\
\sigma_3&=& d\psi + \cos\theta d\phi.\nonumber
\eeqa


If one requires that the spin connection $\omega_{\mu \alpha\beta}$ is
self-dual 
\beqa
\omega_{\mu \alpha\beta}&=&
+\frac12 \epsilon_{\alpha\beta}^{~~~\gamma\delta}\omega_{\mu \gamma\delta},
\eeqa
then $a$, $b$ and $c$ must satisfy
\beqa
\frac{\dot{a}}{abc}&=&+\frac{a^2-b^2-c^2}{2bc},\n
\frac{\dot{b}}{abc}&=&+\frac{b^2-c^2-a^2}{2ca},\\
\frac{\dot{c}}{abc}&=&+\frac{c^2-a^2-b^2}{2ab},\nonumber
\eeqa
where ~$\dot{}=\frac{d}{dt}$.
These conditions are satisfied by the Eguchi-Hanson metric written in the form (\ref{BianchiIXA}).
More generally, if one requires that the Riemann curvature tensor is self-dual:
\beqa
R_{\mu\nu \alpha\beta}&=&
+\frac12 \epsilon_{\alpha\beta}^{~~~\gamma\delta}R_{\mu\nu \gamma\delta},
\eeqa
the conditions are \cite{GM}
\beqa
\frac{\dot{a}}{abc}&=&+\frac{a^2-b^2-c^2}{2bc}+\lambda_a,\n
\frac{\dot{b}}{abc}&=&+\frac{b^2-c^2-a^2}{2ca}+\lambda_b,\label{abcequations}
\\
\frac{\dot{c}}{abc}&=&+\frac{c^2-a^2-b^2}{2ab}+\lambda_c.\nonumber
\eeqa
with
\beqa
\lambda_a=\lambda_b\lambda_c,~
\lambda_b=\lambda_c\lambda_a,~
\lambda_c=\lambda_a\lambda_b.
\label{lambdaconditions}
\eeqa
Since (\ref{lambdaconditions}) is invariant under the sign flips
\beqa
a&\mapsto&+a,\n
b&\mapsto&-b,\n
c&\mapsto&-c,\n
\lambda_a&\mapsto&+\lambda_a,\n
\lambda_b&\mapsto&-\lambda_b,\n
\lambda_c&\mapsto&-\lambda_c
\eeqa
and analogous cyclic generalizations, the consistent values of $\lambda_i$'s are
essentially $(\lambda_a,\lambda_b,\lambda_c)=(0,0,0)$ or $(1,1,1)$ only.
The latter choice corresponds to the cases of the Taub-NUT and the Atiyah-Hitchin 
spaces.

In the original Atiyah-Hitchin paper \cite{AH}, the overall signs of all the 
three equations are flipped, which correspond to an {\it anti}-self-dual curvature 
in our convention. 
See below for more detail on the Atiyah-Hitchin metric in the original 
literature.


\subsection*{A.2~ The Darboux-Halphen system}
This review is based on \cite{Ohyama}.
As was shown in \cite{AH}, by a change of variables
\beqa
w_1=bc,~ w_2=ca,~ w_3=ab, \label{wabc}
\eeqa 
the equations (\ref{abcequations}) with $\lambda_a=\lambda_b=\lambda_c=1$
reduce to
\beqa
\dot{w}_1 + \dot{w}_2 = 2 w_1  w_2, \n
\dot{w}_2 + \dot{w}_3 = 2 w_2  w_3, \label{DH}
\\
\dot{w}_3 + \dot{w}_1 = 2 w_3  w_1, \nonumber  
\eeqa
which is known as the Darboux-Halphen system.
%
As we mentioned in section 2, 
\beqa
w_j(t)&=&2\frac{d}{dt}\log\vartheta_{j+1}(0,i\alpha t)~~~(j=1,2,3),
\label{wj}
\eeqa
($\alpha>0$) is a solution \cite{Ohyama, HP}.
This can be shown by using the composition formula of the 
theta functions and the fact that the theta functions fulfill the 
heat equation.

The original Atiyah-Hitchin solution can be written in this form as
\beqa
w^{AH}_j(\eta)&=&-2\frac{d}{d\eta}\log\vartheta_{j+1}(0,-i\pi\eta)~~~(j=1,2,3),
\label{wAH}
\eeqa
while the solution obtained by Gibbons-Manton \cite{GM} is
\beqa
w^{GM}_j(\eta)&=&-2\frac{d}{d\eta}\log\vartheta_{j+1}(0,-2i\pi\eta)~~~(j=1,2,3).
\label{wGM}
\eeqa
As we remarked at the end of the last subsection, (\ref{wAH}) has 
extra minus signs since they satisfy the anti-self-dual equations (see (\ref{DHminus})).
Comparisons of the solutions are made in 
the next section.

\subsection*{A.3~ Comparison with solutions in the literature}

In this section we derive eqs.(\ref{wAH}) and (\ref{wGM}).

\subsubsection*{A.3.1~ Atiyah-Hitchin's original solution}


Let us first consider  (\ref{wAH}). 
In \cite{AH},  a set of solutions 
are given to the Darboux-Halphen system
\beqa
w'_1 + w'_2 &= &-2 w_1  w_2, \n
w'_2 + w'_3 &=& -2 w_2  w_3, \label{DHminus}
\\
w'_3 + w'_1 &=& -2 w_3  w_1, \nonumber  
\eeqa
which corresponds to $a$, $b$ and $c$ satisfying
\beqa
\frac{a'}{abc}
&=&-\frac{a^2-b^2-c^2}{2bc}-1,\n
\frac{b'}{abc}&=&-\frac{b^2-c^2-a^2}{2ca}-1,\label{ASDabcequations}
\\
\frac{c'}{abc}&=&-\frac{c^2-a^2-b^2}{2ab}-1,\nonumber
\eeqa
where the prime $'$ denotes $\frac d{d\eta}$. 
The metric is assumed to have the form (\ref{BianchiIXA}) with 
the `radial' coordinate $\eta$ replaced instead of $t$. 
The relations between $w_j$'s and $a,b,c$ are (\ref{wabc}).
The solutions are given in terms of the coordinate $\beta$ such 
that
\beqa
\frac{d\beta}{d\eta}&=&u^2,
\label{dbetadeta}
\eeqa
where $u=u(\beta)$ is a function of $\beta$ satisfying
\beqa
\frac{d^2 u}{d\beta^2}+\frac{u^2}{4 \sin^2\beta} =0. 
\label{uequation}
\eeqa
A solution to (\ref{uequation}) is explicitly 
\beqa
u(\beta)&=&(2\sin\beta)^{\frac12} K(\sin\frac\beta2),
\label{u}
\eeqa
where 
\beqa
K(k)&=&\int_0^{\frac\pi 2} \frac{d\phi}{(1-k^2\sin^2\phi)^{\frac12}}
\eeqa
is the complete elliptic integral of the first kind. According to \cite{AH}, we set
\beqa
w_2^{AH}&=&-u\frac {du}{d\beta} + \frac{u^2}{2}\cot\beta,\n
w_3^{AH}&=&-u\frac {du}{d\beta} + \frac{u^2}{2}\frac1{\sin\beta},\\
w_1^{AH}&=&-u\frac {du}{d\beta} - \frac{u^2}{2}\frac1{\sin\beta}.\nonumber
\eeqa 
Using (\ref{u}), we have
\beqa
w_2^{AH}&=&-\frac{d}{d\eta}\log K(\sin\frac\beta 2),\n
w_3^{AH}&=&-\frac{d}{d\eta}\log\left(\sin\frac\beta 2 ~ K(\sin\frac\beta 2)\right),\\
w_1^{AH}&=&-\frac{d}{d\eta}\log\left(\cos\frac\beta 2 ~ K(\sin\frac\beta 2)\right).\nonumber
\eeqa 
The elliptic theta functions and the elliptic integrals are known to be related 
as 
\beqa
K(k)&=&\frac\pi2 \vartheta_3(0,\tau(k))^2,\n
\sqrt{1-k^2} K(k)&=&\frac\pi2 \vartheta_4(0,\tau(k))^2,\\
k K(k)&=&\frac\pi2 \vartheta_2(0,\tau(k))^2,\nonumber
\eeqa
where the modulus $\tau(k)$ is given by
\beqa
\tau(k)=i\frac{K(\sqrt{1-k^2})}{K(k)}. \label{tau}
\eeqa

Let us consider $w_2^{AH}$. Using these equations, we can further rewrite
the expression of $w_2^{AH}$ as
\beqa
w_2^{AH}&=&-2\frac d{d\eta} \log\vartheta_3\left(
0,i
\frac{K(\cos\frac\beta 2)}{K(\sin\frac\beta 2)}
\right).
\eeqa
Using the differential relations among the elliptic functions one can show 
the relation \cite{Ohyama} 
\beqa
\frac{d}{d\beta}\tau({\textstyle\sin\frac\beta 2})
&=&\frac {-i\pi }{2\sin\beta~ K(\sin\frac\beta 2)^2}.\label{dtaudbeta}
\eeqa
The right hand side is equal to $\frac{-i\pi}{u(\beta)^2}$.
Comparing (\ref{dtaudbeta}) with (\ref{dbetadeta}), we find
\beqa
\frac{d\tau({\textstyle \sin\frac\beta 2})}{d\beta}=-i\pi\frac{d\eta}{d\beta}.
\eeqa
If we choose the arbitrary integration constant to be $0$,
we have
\beqa
\tau({\textstyle \sin\frac\beta 2})&=&-\pi i \eta,
\label{tau=ipieta}
\eeqa 
and
\beqa
w_2^{AH}&=&-2\frac{d}{d\eta}\log \vartheta_3(0,-i \pi \eta).
\eeqa
Expressions for $w_3^{AH}$ and $w_1^{AH}$ can be obtained similarly.

A remark is in order here. Since the imaginary part of the elliptic modulus of 
must be positive, (\ref{tau=ipieta}) implies that $\eta$ takes its value from 
$-\infty$ to $0$, which is puzzling if it is regarded as a radial coordinate.
Since the system of equations (\ref{DHminus}) is invariant under the simultaneous 
sign flips  
$w_j\mapsto -w_j$ and $\eta\mapsto -\eta$, one might wonder if
\beqa
w_j^{AH}&\stackrel{?}{=}&-2\frac{d}{d\eta}\log \vartheta_{j+1}(0,+i \pi \eta)
\label{+?} 
\eeqa
also work. They do not, however, because, although (\ref{+?}) still solves 
(\ref{DHminus}),  $a^2$, $b^2$ and $c^2$ derived from (\ref{+?}) 
with (\ref{wabc}) turn out to be {\it negative}. This problem can be cured by either 
changing the relations (\ref{wabc}) to ones with  minus sign: $w_1=-bc,$ etc.,  
or replacing  (\ref{DHminus})  with the self-dual equations  (\ref{DH}) without 
minus signs on the right hand sides; in any case one is forced to modify some 
of the assumptions made in \cite{AH}. We have numerically checked that (\ref{wAH}) 
(and not (\ref{+?}))
is correct by Mathematica.

While, of course, these complications are not essential, we adopted (\ref{DH}) 
instead of (\ref{DHminus}) in the text for a simpler notation with less minus signs.
Note that, unlike $\eta$,  the coordinate $t$ in (\ref{wj}) takes its value from $0$ 
to $+\infty$.

\subsubsection*{A.3.2~ Gibbons-Manton's solution}

In \cite{GM}, the radial coordinate was chosen so that 
the Bianchi IX metric took the following form:
\newcommand\rt{\tilde{r}}
\beqa
ds^2&=&\frac{b^2}{\rt^2}d\rt^2 + a^2\sigma_1^2
+b^2 \sigma_2^2 
+c^2 \sigma_3^2.
\label{GMBianchiIX}
\eeqa
$a$, $b$ and $c$ which solve 
\beqa
-\frac\rt b \frac{da}{d\rt}
&=&-\frac{a^2-b^2-c^2}{2bc}-1,\n
-\frac\rt b \frac{db}{d\rt}&=&-\frac{b^2-c^2-a^2}{2ca}-1,\label{GMabcequations}
\\
-\frac\rt b \frac{dc}{d\rt}&=&-\frac{c^2-a^2-b^2}{2ab}-1\nonumber
\eeqa
give an anti-self-dual metric. The solution was similarly given in terms of 
$w_j=w^{GM}_j$ ($j=1,2,3$) (\ref{wabc}) as
\beqa
w_2^{GM}&=&- \rt \sin\beta\frac {d\rt}{d\beta},\n
w_3^{GM}&=&- \rt \sin\beta\frac {d\rt}{d\beta}+\frac{\rt^2}2(1-\cos\beta),\\
w_1^{GM}&=&- \rt \sin\beta\frac {d\rt}{d\beta}-\frac{\rt^2}2(1-\cos\beta)\nonumber
\eeqa 
with
\beqa
\rt&=&2K\left(
\sin\frac\beta 2
\right).\label{rtilde}
\eeqa
To derive (\ref{wGM}), we change variables so that the metric 
takes the form  (\ref{BianchiIXA}):
\beqa
abc \;d\tilde{\eta} = -\frac b\rt d\rt.
\eeqa
$\tilde{\eta}$ is the new radial coordinate. Then (\ref{rtilde}) implies that 
\beqa
\frac{d\tilde{\eta}}{d\beta}&=&\frac1{2u(\beta)^2},
\label{detatildedbeta}
\eeqa
where $u(\beta)$ is the function (\ref{u}). 
On the other hand, $w^{GM}_2$ can be similarly written as
\beqa
w^{GM}_2&=&-\frac d{d\tilde\eta} \log K\left(
\sin\frac\beta 2
\right)\\
&=&
-2\frac d{d\tilde\eta} \log\vartheta_3\left(
0,i
\frac{K(\cos\frac\beta 2)}{K(\sin\frac\beta 2)}
\right).
\eeqa
Comparing (\ref{dtaudbeta}) with (\ref{detatildedbeta}), we have in this case
\beqa
\tau\left(
\sin\frac\beta 2
\right)&=& -2\pi i \tilde \eta,
\eeqa
and therefore
\beqa
w_2^{GM}&=&-2\frac{d}{d\tilde\eta}\log \vartheta_3(0,-2 i \pi \tilde\eta).
\eeqa
The proofs for $w_3^{GM}$ and $w_1^{GM}$ are also straightforward.

We have shown that the value of $\alpha$ (\ref{wj}) for Gibbons-Manton's solution 
is $2\pi$.
Plugging $\alpha=2\pi$ into (\ref{negativeT-NUT}), 
we find agreement 
with the asymptotic metric obtained in \cite{GM}.

\section*{Acknowledgments}
We thank M.~Tanaka, S.~Tomizawa and Y.~Yoshida for discussions and comments.
The work of S.~M. is supported by 
Grant-in-Aid
for Scientific Research (C) \#20540287 and also by 
(A) \#22244030 
from
the Ministry of Education, Culture, Sports, Science
and Technology of Japan.
%


\end{document}